# Muon spin rotation study of magnetism and superconductivity in $BaFe_{2-x}Co_xAs_2$ and $Pr_{1-x}Sr_xFeAsO$

C. Bernhard[1], A. J. Drew[1,2], L. Schulz[1], V.K. Malik[1], M. Rössle[1], Ch. Niedermayer[3], Th. Wolf[4], G.D. Varma[5], G. Mu[6], H.-H. Wen[6], H. Liu[7], G. Wu[7], and X.H. Chen[7]

[1]Department of Physics and Fribourg Center for Nanomaterials - Frimat, University of Fribourg, Chemin du Musée 3, CH-1700 Fribourg, Switzerland
[2]Queen Mary University of London, Mile End Road, London E1 4NS, UK
[3]Laboratory for Neutron Scattering, ETHZ & PSI, CH-5232 Villigen PSI, Switzerland
[4]Forschungszentrum Karlsruhe, IFP, D-76021 Karlsruhe, Germany
[5]Department of Physics, Indian Institute of Technology Roorkee, Roorkee 247667, India
[6]National Laboratory for Superconductivity, Institute of Physics and Beijing National Laboratory for Condensed Matter Physics, Chinese Academy of Science, P.O. Box 603, Beijing 100190, Peoples Republic of China
[7]University of Science and Technology of China, Hefei, Anhui 230026, Peoples Republic of China

We present a muon spin rotation (μSR) study of the magnetic and superconducting properties of single crystals of electron-doped $BaFe_{2-x}Co_xAs_2$ with x=0.08, 0.20, and 0.25 ($T_c$=9, 25 and 20K) and of polycrystalline hole-doped $Pr_{1-x}Sr_xFeAsO$ with x=0 and 0.2 ($T_c$=15 K). In the former series we observe some interesting parallels with the electron doped $SmFeAsO_{1-x}F_x$ 1111-type system [A.J. Drew et al., to appear in Nature Materials 2009 and arXiv:0807.4876]. In particular, we obtain evidence that strongly disordered static magnetism coexists with superconductivity on a microscopic scale in underdoped samples and even at optimum doping there is a slowing down (or enhancement) of dynamic magnetic correlations below $T_c$≈25K. To the contrary, for the hole-doped $Pr_{1-x}Sr_xFeAsO$ samples we obtain evidence for a mesoscopic phase segregation into regions with nearly unperturbed AF order and others that are non magnetic and most likely superconducting. The observed trend resembles the one that was previously reported for hole-doped $Ba_{1-x}K_xFe_2As_2$ [A.A. Aczel et al., Phys. Rev. **B 78**, 214503 (2008); J.T. Park et al., arXiv:0811.2224] and thus might be a common property of the hole doped systems.

**Introduction**

The recent discovery of high temperature superconductivity (HTSC) with values of the critical temperature, $T_c$, up to 55 K in the quaternary oxypnictide $RFeAsO_{1-x}F_x$ (so-called 1111-compound) where R is a lanthanide, has attracted a great deal of attention [1-3]. This is the highest $T_c$ value that has been observed in a copper-free superconductor which has a layered structure and likely has correlated charge carriers similar to the well known cuprate HTSC. Soon after, the oxygen-free pnictide superconductors $(Ba,Sr)_{1-x}K_xFe_2As_2$ and $(Ba,Sr)Fe_{2-x}Co_xAs_2$ with $T_c$ up to 38 K and 25 K [4-7] (so-called 122-compounds) have been discovered.

Similar to the cuprates, the superconductivity in the pnictides emerges upon doping of a parent material that exhibits long-range antiferromagnetic (AF) order [8-10]. These materials are therefore prime candidates for being unconventional superconductors where the pairing mechanism is based on electronic or magnetic correlations rather than on the conventional electron-phonon interaction.

Nevertheless, these pnictides also provide several unique aspects which clearly distinguish them from the cuprate HTSC. For example, the undoped parent compounds, such as LaFeAsO or $(Ba,Sr)Fe_2As_2$, are metals (even though fairly poor ones) as opposed to the cuprates, such as $La_2CuO_4$ or $YBa_2Cu_3O_6$, which are so-called Mott-Hubbard insulators where the strongly repulsive electronic correlations give rise to an insulating ground state despite a half-filled conduction band. Furthermore, band structure calculations suggest that several (probably all five) of the Fe(3d)-As(4p) subbands are crossing the Fermi-level and thus participate in the metallic response [11-14]. This is in contrast to the cuprates, where due to a strong Jahn-Teller distortion of the $CuO_6$ octahedra only the Cu($3d_{x^2-y^2}$)-O($2p_{x,y}$) band is relevant. This opens up various interesting aspects like multiband superconductivity with different possibility of mixing of the SC order parameters and a possible relevance of strong orbital fluctuations possibly contributing to the pairing. Besides the likely dominant Fe(3d)-As(4p) hybridization, there also appears to be a sizeable direct Fe(3d)-Fe(3d) overlap [11] which opens up various possibilities for competing kinds of exchange interactions and subsequent magnetic frustration and consequent low-energy excitations which may be relevant for the superconducting pairing interaction.

It is therefore of great importance to obtain further insight into the differences and similarities of the pnictide and cuprate HTSC. One of the prominent issues concerns the question of how magnetism and superconductivity evolve upon carrier doping in these pnictides and whether they coexist.

In this context it is important to specify the differences of the electronic doping in the pnictides. Unlike the cuprates, the undoped parent compounds in the pnictides are already metals, even though fairly bad ones. Their Fermi surface consists of small hole-like and electron-like pockets that are of similar size and shape located at different positions in the Brillouin-zone (at the center and the boundary of the BZ, respectively). The resulting nesting condition is believed to be at the heart of the formation of the commensurate spin density wave (SDW) state which can also be viewed as itinerant antiferromagnet (AF) state. Electron doping, for example due to Co-substitution in $BaFe_{2-x}Co_xAs_2$, raises the Fermi-level and thus reduces the size of the hole pockets and increases the size of the electron pockets. It thus weakens the nesting condition and gives rise to a suppression of the SDW state and subsequently to the occurrence of superconductivity. Unlike in the cuprates, the superconducting state is composed not only of the doped carriers but also of the carriers that are present already in the undoped compound.

Furthermore, it should be noted that the electronic structure as derived from band structure calculations is strongly asymmetric with respect the electron and hole doping [12, 13]. While the density of states is rather low and featureless towards the electron doped side, it exhibits a rather steep increase on the hole doped side which suggests that some flatter parts of the hole like bands are approaching the Fermi surface. Accordingly, it can be expected that the phase diagram as a function of the hole- and electron doping is fairly different.

For the 1111-compounds it has recently been shown that static magnetic order persists upon doping, at least up to the point where SC first emerges. For La-1111 it was found that an abrupt first order like phase transition occurs between a static magnetic state at x≤0.05 and a non-magnetic superconducting one at x≥0.05 [15], albeit a recent similar μSR study observes static magnetism even at x=0.06 [16]. For Sm-1111 it was shown that static magnetism persists to higher F-doping and coexists with superconductivity over a significant range of the phase diagram [17]. In addition, slow magnetic fluctuations were shown to coexist with superconductivity even beyond optimum doping where the highest $T_c$ value is reached [18]. It was also found that the transition from a tetragonal to orthorhombic structure at low temperature [19] tracks the transition temperature of the static magnetism [17]. This observation agrees with previous reports that the structural transition is a prerequisite for the magnetic one [8, 9]. For Ce-1111 a magnetic phase diagram similar to La-1111 has been reported based on neutron scattering measurements [20] which implied that static magnetism and superconductivity do not coexist. Nevertheless, the structural data yielded a doping dependence of the tetragonal to orthorhombic phase transition line similar to the one in Sm-

1111 [19, 20]. Furthermore, recent µSR measurements on Ce-1111 indicate that short range magnetic order (for which the corresponding Bragg peaks were not resolved in the neutron diffraction experiments) persists well into the superconducting regime [16, 21]. It appears that the doping dependence of short-ranged static magnetism and superconductivity depends rather crucially on the ionic radius (and the subsequent chemical pressure) and/or the magnetic moment of the R-site ion. Notably, the highest $T_c$ values occur for the systems where static magnetism persists to higher doping and thus coexists with superconductivity. This has lead to speculations that soft magnetic fluctuations may be very beneficial for superconductivity [17, 18].

For the 122-compound, where sizeable single crystals are now available, the µSR studies have been focused on the hole doped system $(Ba,Sr)_{1-x}K_xFe_2As_2$ [21-23]. Here it was also observed that static magnetism and superconductivity coexist up to doping levels of at least x=0.35 [21-23]. However, different from Sm-1111 where the magnetic volume fraction is close to 100% [17], for $(Ba,Sr)_{1-x}K_xFe_2As_2$ the µSR experiment provide evidence for a mesoscopic phase separation. In the latter µSR resolves two different fractions that arise from non superconducting regions where AF order is maintained similar as in the undoped compound and others that are entirely non magnetic and superconducting [21-23]. The observed behaviour is reminiscent of the so-called super-oxygenated cuprate compound $La_2CuO_{4+\delta}$ where a macroscopic phase separation into hole poor regions with a fully developed AF (or SDW) order and hole rich and superconducting regions is well established [24-27]. For $La_2CuO_{4+\delta}$ it was shown that this mesoscopic electronic phase segregation is linked to the formation of a superstructure that involves a clustering of the excess oxygen ions which helps to circumvent the strong Coulomb repulsion that would counteract a purely electronic phase segregation [24-27]. In comparison, to date there exist no experimental evidence for a similar clustering of the $K^+$ ions in $(Ba,Sr)_{1-x}K_xFe_2As_2$ [21-23]. Alternatively, the spatial variation in the electromagnetic properties may not even be caused by a modulation of the carrier density in the FeAs layers. This possibility is suggested by recent reports that AF order can be suppressed and superconductivity induced simply by applying pressure to the undoped compound $BaFe_2As_2$ [28, 29] or by isoelectronic substitution of P for As in $BaFe_2As_{1-x}P_x$ [30] both of which do not involve a carrier doping effect but may well influence the balance between competing exchange interactions or affect the nesting condition of the band structure.

In this context, it is interesting to learn how magnetism and superconductivity evolve in electron doped 122-systems and hole doped 1111-compounds. In the following we address this question by presenting first µSR data on single crystals of $BaFe_{2-x}Co_xAs_2$ with 0≤x≤0.25 where Co gives rise to electron doping [31-33] and on polycrystalline samples of the hole doped 1111-compound $Pr_{1-x}Sr_xFeAsO$ [34, 35].

## Sample Preparation and Characterisation

The $BaFe_{2-x}Co_xAs_2$ single crystals were grown from self-flux with a technique similar to the one described in Ref. [31]. Their Co content was determined by energy dispersive x-ray (EDX) analysis. Their crystal structure and purity was checked with x-ray analysis. The superconducting critical temperatures, $T_c$, have been deduced from the resistivity and dc magnetisation data that are shown in Fig. 1. The onset of the drop in resistivity in Fig. 1(a) marks an onset superconducting transition temperature $T_c^{ons}$=9, 25 and 20 K at x=0.08, 0.2 and 0.25, respectively. The diamagnetic signal of the susceptibility data in Fig. 1(b) and the drop to zero resistivity occur at somewhat lower temperatures of 6, 23 and 19 K, respectively. These values agree with the ones in Ref. [32] but not with Ref. [33] where a maximum $T_c$ value of 22 K is reported at a significantly lower doping of x≈0.12. Presently we do not know whether this discrepancy is related to a difference in the growth conditions or whether it rather reflects the uncertainty in determining the absolute Co contents.

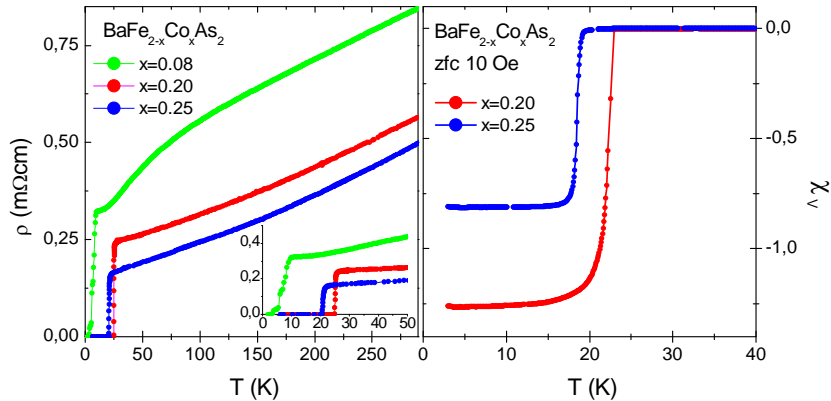

Figure 1: **(a)** T-dependent in-plane resistivity of the $BaFe_{2-x}Co_xAs_2$ single crystals with x=0.08, 0.2 and 0.25, respectively. **(b)** Volume susceptibility as deduced from zero-field cooled measurements in a field of 10Oe applied parallel to the c-axis of the crystals.

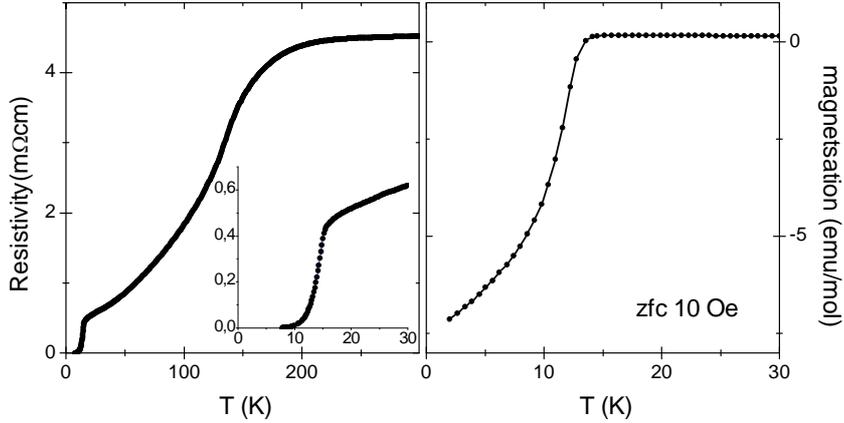

Figure 2: T-dependent resistivity and dc zero-field cooled (zfc) magnetisation of a polycrystalline $Pr_{0.8}Sr_{0.2}FeAsO$ sample with $T_c^{ons} \approx 15$ K.

The polycrystalline $Pr_{1-x}Sr_xFeAsO$ samples with x=0 and 0.2 where grown with a solid state reaction method as described in Ref. [35]. Standard x-ray diffraction measurements confirmed that the samples consisted predominantly of the 1111-phase. For the Sr-doped sample they also revealed the presence of a minority FeAs phase [35]. As shown in Fig. 2, the resistivity and susceptibility measurements yield a $T_c$ value of about 15 K.

## Muon Spin Rotation

The muon spin rotation (µSR) experiments have been performed at the GPS instrument of the πM3 beamline at Paul Scherrer Institut (PSI) in Villigen, Switzerland with 100% spin polarized, positive "surface muons". The µSR technique is especially suited for the study of magnetic and superconducting materials since it allows one to investigate the local magnetic field distribution on a microscopic scale and to directly access the corresponding volume fractions [36]. For example in strongly underdoped cuprate HTSC [37-40] or in the ruthenate-cuprate HTSC [41] it has been very successful in revealing the coexistence of magnetism and superconductivity. The µSR technique typically covers a time window of $10^{-6}$ to $10^{-9}$ seconds and allows one to detect internal magnetic fields between 0.1G and several Tesla. The muon beam is implanted into the bulk of the sample where they thermalize very rapidly ($\sim 10^{-12}$s) without any noticeable loss in their initial spin polarization. Each muon stops at a well-defined interstitial lattice site. For the pnictides these muon sites have not yet been accurately determined. However, in analogy to the cuprates they are likely bound to the negatively charged O or As ions. The ensemble of muons is randomly distributed throughout a layer of 100-200 μm thickness and therefore probes a representative part of the sample volume.

For the mosaics of thin $BaFe_{2-x}Co_xAs_2$ single crystals with x=0.2 and 0.25 we used a thin (100 micrometer) Ag foil in front of the sample (known as a "degrader") to reduce the energy of the incoming muons and thereby to increase the number of muon that stop in the sample. To suppress the signal from the muons that missed the sample or passed through it, we used a veto-counter that is situated behind the sample. The sample and the degrader were mounted on an open frame sample holder with a very thin mylar foil that was transparent to the muon beam.

Each muon spin precesses in its local magnetic field $B_\mu$ with a precession frequency, $\nu_\mu = \gamma_\mu/2\pi \cdot B_\mu$, where $\gamma_\mu = 2\pi \cdot 135.5$ MHz/T is the gyromagnetic ratio of the positive muon. Positive muons decay with a mean life time of $\tau_{\mu^+} \approx 2.2 \mu s$ into two neutrinos and a positron which is preferentially emitted along the direction of the muon spin at the instant of decay. The time evolution of the spin polarization P(t) of the muon ensemble can therefore be obtained via the time-resolved detection of the spatial asymmetry, A(t), of the decay positron emission rate. More details regarding the μSR technique can be found in Ref. [36-41].

## μSR experiments on electron-doped $BaFe_{2-x}Co_xAs_2$

Figure 3 shows so-called transverse field (TF) μSR data where an external magnetic field was applied in the direction perpendicular (transversal) to the initial muon spin direction. This TF was directed perpendicular to the $Fe_2As_2$ layers. Figure 3(a) shows the result of a so-called vortex-lattice pinning experiment. The crystals were first field cooled here in a TF field of 0.1 Tesla to 5 K and thus well into the superconducting state where a vortex lattice develops. Subsequently, the magnetic field was reduced by 0.005 Tesla before the μSR measurement was started. The resulting distribution of the magnetic fields at the muon sites, the so-called μSR lineshape, as obtained with a maximum entropy analysis [42] is shown in Fig. 3(a) for the crystals with x=0.2 and x=0.25. In both cases we obtained two well separated peaks that are centred around 0.095 and 0.1 Tesla, respectively. The former peak arises from the background of muons that missed the sample and stopped in the degrader, the windows of the cryostat or the sample holder. The peak at 0.1 Tesla is due to the muons that stop in the sample where a vortex lattice forms below $T_c$ that is strongly pinned by defects such that the magnetisation density cannot follow the change of the external magnetic field similar as it was previously observed for the cuprate HTSC [43]. The observed splitting between the position of the background signal and the vortex lattice signal and the large amplitude of the latter (despite of the small volume of these crystals) is a clear indication that our crystals are bulk superconductors. In addition, it provides us with an elegant way of separating the background signal from the one of the sample and thus to determine the corresponding amplitudes.

We also performed corresponding TF-μSR measurements where the sample was field cooled in 0.1 Tesla without any subsequent changes. Here the background signal was superimposed on the one due to the vortex lattice but has been subtracted using the parameters as obtained from the pinning experiments. This resulted in a vortex lattice lineshape at 5 K with a second moment, $<\Delta B^2>^{1/2}$, of about 2.3 and 1.8 mT at x=0.2 and 0.25, respectively. Assuming the London limit ($\lambda >> \xi$), a triangular symmetry of the vortex lattice, and neglecting the influence of pinning induced disorder (which is certainly sizeable here) this second moment is related to the in-plane magnetic penetration depth, $\lambda_{ab}$, according to [44] $\left\langle \Delta B^2 \right\rangle^{1/2} = \left( \frac{0.00371 \Phi_0^2}{\lambda_{ab}^4} \right)^{1/2}$, where $\Phi_0$ is the magnetic flux quantum. Thus we arrive at lower limits of the in-plane magnetic penetration depth of $\lambda_{ab} \geq 230$ and 275 nm for x=0.2 and x=0.25, respectively. On a so-called "Uemura-plot" [45] these samples thus fall close to the line that is common to the underdoped cuprate HTSC, similar as it was previously reported for electron doped La- and Sm-1111 [15, 18]. However, one must be careful with this type of analysis as there are a number of systematic errors. In high κ materials in small and intermediate magnetic fields, the small coherence length and diverging magnetic fields close to the core mean that the high-field tail of the vortex lineshape has a small amplitude which makes it difficult to distinguish between signal and noise. As a consequence, the penetration depth can be systematically overestimated. Furthermore and perhaps more important, any disorder present in the vortex lattice can either lead to an over- or underestimation of the penetration depth, depending on the exact form of disorder present (e.g static line disorder [46] or flexible vortices [47]). In our case, it is likely that static line disorder due to pinning will be the most important factor in determining the error, leading to lineshape broadening and therefore an under-estimate in the penetration depth. In combination with the uncertainty related to the unknown vortex lattice symmetry and systematic errors in subtracting the relatively large non-superconducting background, the quoted values of the penetration depths should thus be considered as a first estimate rather than the definitive measurement.

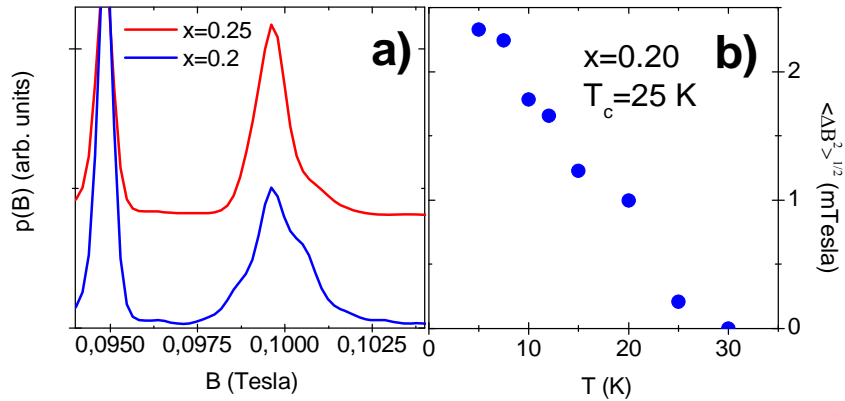

Figure 3: TF-µSR data on a set of small single crystals of $BaFe_{2-x}Co_xAs_2$ with x=0.2 and 0.25 and $T_c^{ons}$= 25 and 20 K, respectively. **(a)** TF-lineshapes (obtained from a maximum entropy analysis) for field-cooling in 0.1 Tesla to 5 K with subsequent field reduction by 0.005 Tesla. Due to the pinning of the vortex lattice, the signal from the muons that stop inside the samples remains centered around 0.1 Tesla (pinning effect of the vortex lattice). To the contrary, the signal from the muons that stop outside the sample (background signal) occurs at the external field of 0.095 Tesla. **(b)** *T*-dependence of the second moment of the µSR lineshape, $<\Delta B^2>^{1/2}$, of the x=0.2 sample.

Figure 3(b) shows the T-dependence of $<\Delta B^2>^{1/2}$ for the x=0.2 sample which confirms that the lineshape broadening sets in below $T_c$=25 K and thus is mainly caused by the formation of the vortex lattice. Nevertheless, given the uncertainties as mentioned above and the indications that magnetic correlations provide an additional contribution (see below), we refrain here from further discussing the T-dependence of $\lambda_{ab}(T)$ and its possible implications on the symmetry of the superconducting order parameter.

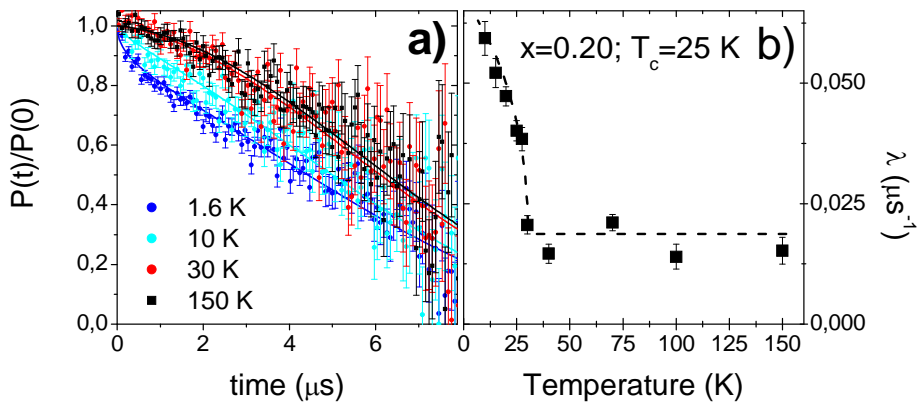

Figure 4: ZF-data on the optimally doped BaFe$_{1.8}$Co$_{0.2}$As$_2$ single crystals with T$_{sc}$=25 K. **(a)** Time-resolved spectra at representative temperatures, and **(b)** *T*-dependence of the relaxation rate, λ, which exhibits a noticeable increase below T$_{sc}$=25 K.

Instead we move on to the discussion of the ZF-µSR data in the optimally doped x=0.20 sample which reveal the existence of weak magnetic correlations that set in right below T$_c$=25 K. Figure 4(a) shows representative ZF-µSR time spectra at different temperatures. The solid lines show fits that were obtained with the relaxation function:

$$A(t) = A \cdot GKT \cdot \exp(-\lambda t)^\beta + A_{BG}(0).$$

The Gaussian Kubo-Toyabe (GKT) function accounts for randomly oriented, paramagnetic moments that are most likely of nuclear origin [36]. Its relaxation rate Δ=0.12µs$^{-1}$ was determined at 150K and fixed for fitting the spectra at lower temperature. The so-called stretched exponential function with the relaxation rate, λ, and exponent, β, accounts for the contribution of the electronic magnetic moments. The exponent β was found to be close to 1 except at T ≤ 5 K where it decreased to a value of about 0.5. Such a behavior is suggestive of a slowing down of dynamic spin correlations whose fluctuation rate becomes spatially inhomogeneous [36]. Additional longitudinal-field (LF) µSR measurements performed at 5 K (not shown) confirm that these magnetic correlations remain at least partially dynamic similar as it was previously observed in optimally doped SmFeAsO$_{1-x}$F$_x$ [16]. Figure 4(b) details the temperature dependence of λ which can be seen to increase rather suddenly below T$_c$=25 K. In this context it is interesting to note that a superconductivity-induced change in the spin dynamics in near optimally doped BaFe$_{2-x}$Co$_x$As$_2$ single crystals has also been observed by recent inelastic neutron scattering measurements [48]. In that study a magnetic resonance mode with a maximum around 9 meV and a gap-like suppression at lower energies were found to develop right below T$_c$. Our µSR data reveal that the spin fluctuation spectrum exhibits some corresponding changes in the quasi-static regime. The obtained evidence for a spatial inhomogeneous distribution of the fluctuation time scales may be related for example to the disorder from the Co dopands that reside within the FeAs layers.

Alternatively, our ZF-µSR data could also be described in terms of a macroscopically inhomogeneous scenario of regions that are entirely non magnetic and others that are strongly magnetic. For example, a reasonable fit to the spectrum at low temperature can be obtained with a function that consists of the sum of a pure GKT function with an amplitude of about 70% of the signal and a strongly damped exponential function that accounts for the remaining signal (not shown). Such a macroscopically inhomogeneous state could arise from a gradient in the concentration of the Co ions which could leave parts of the sample in an underdoped state with static magnetic correlations. Likewise, it could be due to inclusions of a magnetic impurity phase like FeAs, albeit such secondary phases do not show up in the x-ray data (not shown). In order to finally differentiate between these different scenarios more detailed μSR measurements on larger crystals will be required. Here we only remark that within the macroscopically inhomogeneous scenario it is not obvious why the relatively sharp onset of the magnetic signal occurs so close to the superconducting transition at $T_c$=25 K [49]. The latter observation rather is suggestive of an intimate relationship between the quasi-static magnetic correlations and superconductivity.

Next we discuss the μSR data on a fairly large underdoped $BaFe_{1.82}Co_{0.08}As_2$ single crystal with $T_c$=9 K which reveal a strongly disordered static magnetism that develops well above $T_c$. Once more the trend is similar as previously reported for underdoped Sm-1111 [17]. Figure 5(a) shows the ZF-μSR time spectra (symbols) at several representative temperatures. The solid lines show fits that were obtained with the function:

$$A(t) = A_1 \cdot \exp(-\lambda_1 t) + A_2 \cdot \cos(\gamma_\mu B_\mu t) \cdot \exp(-\lambda_2 t) + A_3 .$$

The temperature dependences of the obtained relaxation rates $\lambda_1$ and $\lambda_2$ and of the precession frequency $\nu_\mu$, are show in Figs. 5(b) and 5(c), respectively. The three different contributions to the μSR signal can be clearly distinguished at T≤12 K. About 60% of the signal ($A_1$) exhibits a rapid exponential relaxation. The low temperature value of $\lambda_1 \approx 30$ μs$^{-1}$ indicates the presence of sizeable static moments that are strongly inhomogeneous either in their magnitude and/or in their relative orientation. Another 20% of the signal ($A_2$) oscillates, albeit with a fairly large relaxation rate $\lambda_2$. Notably, $\lambda_2$ seems to have a maximum around $T_c^{ons}$=9 K which seems to suggest that the onset of superconductvitiy has a clear impact on the magnetic order. The precession frequency, $\nu_\mu$, reaches a low temperature value of 14 MHz which is about half the one reported for undoped $BaFe_2As$ [22]. The remaining 20% fraction of the signal is non-relaxing. It could arise either from a corresponding non magnetic volume fraction or else it could correspond to the non-relaxing tail of the signal from an fully magnetic sample. Our additional weak transverse field (wTF) measurements (not shown here) support the latter possibility.

Once more, these ZF-spectra could also be described with a different fitting function in terms of a so-called Bessel-function which describes the response of a incommensurate SDW state. Irrespective of this open question we can conclude that this underdoped $BaFe_{1.92}Co_{0.08}As_2$ crystal exhibits bulk but strongly disordered magnetism which becomes static below 15 to 20 K. Notably, the maxmimum in the relaxtion rate around $T_c^{ons}$=9 K is suggestive of a direct interaction between static magnetism and superconductivity. Slow magnetic fluctuations are observed to higher temperatures of at least 25 K

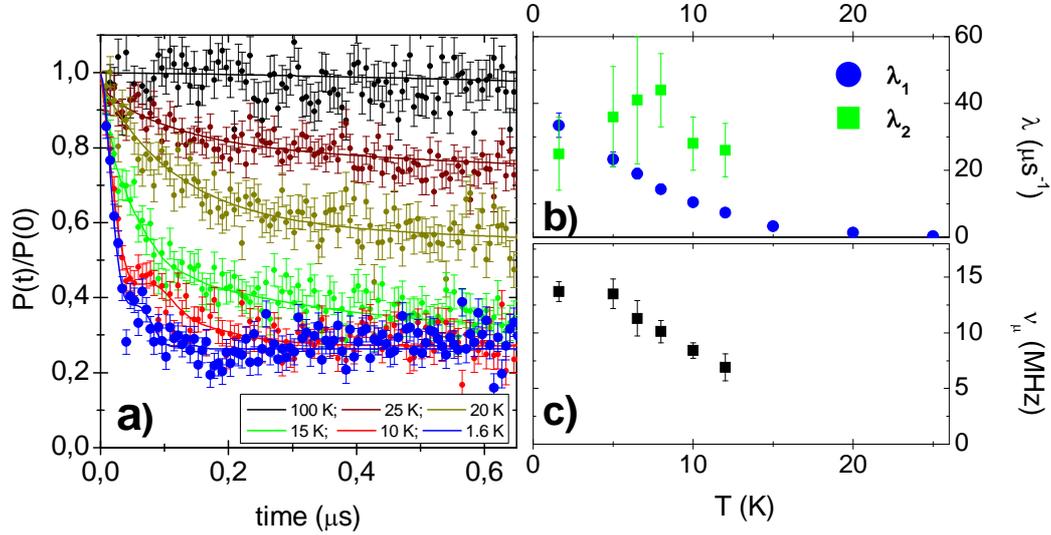

Figure 5: ZF-data of an underdoped $BaFe_{1.92}Co_{0.08}As_2$ single crystal with $T_c^{ons}$=9 K. **(a)** Time-resolved spectra at representative temperatures. **(b)** $T$-dependence of the relaxation rates $\lambda_1$ and $\lambda_2$ and **(c)** of the muon precession frequency $\nu_\mu$.

From our µSR measurements on this crystal we could not obtain any direct evidence of the superconducting state. In the TF-µSR measurements we could not resolve an enhanced relaxation due to the formation of a vortex lattice (due to the magnetism the relaxation is already very high at $T>T_c$), nor could we observe a vortex pinning effect as in the optimal and overdoped samples.

To summarize this section, we observed that the magnetic and superconducting properties of single crystals of the electron-doped 122-compound $BaFe_{2-x}Co_xAs_2$ reveal some interesting analogies with the ones of the electron-doped 1111-compound $SmFeAsO_{1-x}F_x$ [17, 18]. In both cases we find that: (i) bulk, static magnetism persists to a fairly high doping level and even coexists with superconductivity, (ii) slow magnetic fluctuations (on the time scale of the µSR technique) are present even in optimally doped samples where they develop or slow down significantly in the superconducting state. These optimally doped samples are at the same time found to be bulk superconductors whose vortex lattice extends throughout the entire sample volume [17]. Our results also reveal some important differences between the electron-doped and the hole-doped 122-systems. Unlike in hole-doped $(Ba,Sr)_{1-x}K_xFe_2As_2$, our electron-doped BaFe$_{2-x}$Co$_x$As$_2$ crystals show no signs of a mesoscopic phase segregation into nearly undoped and thus magnetically ordered regions and highly doped ones that are entirely non-magnetic and superconducting. This raises the question whether the mesoscopic phase segregation is specific to the (Ba,Sr)$_{1-x}$K$_x$Fe$_2$As$_2$ system.

## µSR experiments on hole-doped Pr$_{1-x}$Sr$_x$FeAsO

In order to further address this challenging question, we have investigated polycrystalline samples of hole-doped Pr$_{1-x}$Sr$_x$FeAsO with x=0 and 0.2. Our ZF- and weak transverse field (wTF) µSR data are summarized in Fig. 6. Representative ZF-µSR time spectra for the undoped sample with x=0 which is in a bulk AF state similar to RFeAsO [1-3, 8-10, 50], where R is a lanthanide ion, are shown in Fig. 6(a). The slowly relaxing spectrum at 160 K is well described by a Gaussian Kube-Toyabe relaxation function (solid line) which accounts for the contribution of nuclear moments. The spectrum at 50 K contains a predominant oscillatory signal with a precession frequency of about 23 MHz (corresponding to an average local field at the muon site of $<B_\mu>$=0.17 Tesla) and a moderate relaxation rate of $\lambda^{ZF}\approx3.5\mu s^{-1}$. The T-dependence of the ZF-µSR precession frequency, $\nu_\mu$, as shown in Fig. 6(c), yields a magnetic transition temperature of the moments in the FeAs layers of $T_N\approx145$ K. The decrease in $\nu_\mu$ below 20 K is most likely due to the additional ordering of the Pr moments. Most important is the large amplitude of the oscillatory signal of about 2/3 of the total µSR signal which suggests that essentially the entire sample volume is magnetically ordered. Note that for a polycrystalline magnetic sample, where the local magnetic field is randomly oriented with respect to the muon spin direction, only 2/3 of the µSR signal oscillates [37-41]. The absence of a sizeable non-magnetic volume fraction is confirmed by the wTF spectrum in Fig. 6(d) which contains only a very small component with a precession frequency of 0.67 MHz (corresponding to the applied field of 50 Oe) that is well accounted for by the background muons which amount to about 5%. Our ZF- and wTF-µSR data thus establish that the undoped PrFeAsO sample is a bulk AF with a transition temperature of $T_N\approx145$ K.

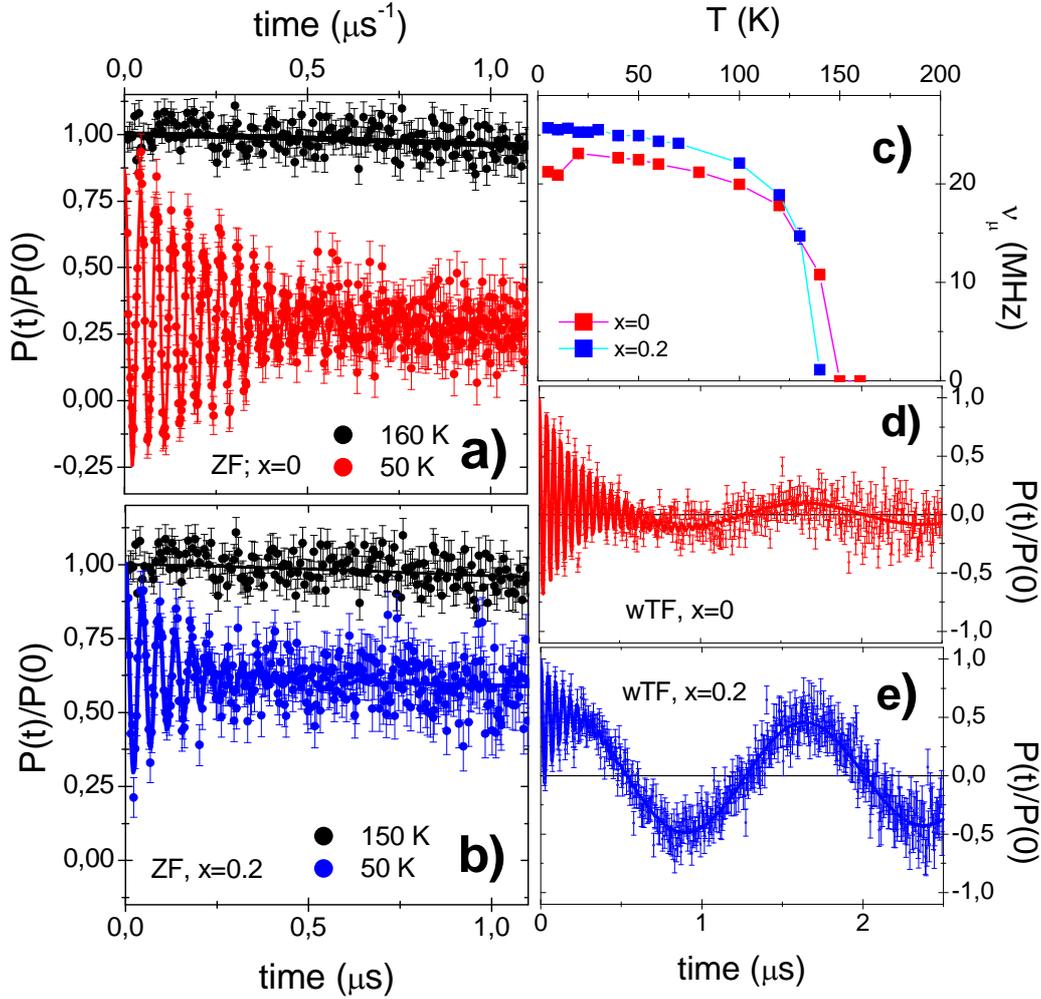

Figure 6: μSR data on polycrystalline samples of hole-doped $Pr_{1-x}Sr_xFeAsO$ with x=0 and x=0.2 ($T_c^{ons}$=15K). **(a)** ZF-μSR spectra at 160 and 50 K of the undoped sample with x=0. **(b)** Corresponding ZF-μSR spectra at T=150 and 50 K of the superconducting sample with x=0.2. **(c)** T-dependence of the precession frequency, $\nu_\mu$, for x=0 and 0.2, and **(d)** and **(e)** weak transverse field data at 50 K for x=0 and x=0.2, respectively.

To our surprise, the corresponding ZF- and wTF-μSR data at x=0.2 reveal that a similar magnetic order persists in the FeAs layers of this hole doped superconducting sample. This is evident from the sizeable oscillatory signal in the low temperature ZF-μSR spectrum in Fig. 6(b) which has a precession frequency of about 25 MHz that is even slightly higher than the one at x=0. Furthermore, the T-dependence of the ZF-μSR precession frequency, $\nu_\mu$, in Fig. 6(c) indicates a magnetic transition temperature of $T_N \approx 130$ K that is only moderately suppressed with respect to the undoped sample. The main difference concerns the enhanced relaxation rate of $\lambda \approx 7\ \mu s^{-1}$ and the reduced amplitude of the oscillatory signal which suggests that about 60% of the sample volume remain magnetic at x=0.2. This estimate is confirmed by the wTF-μSR data in Fig. 6(e) which yield a 40 % fraction (after subtraction of the background signal) that oscillates with a frequency of 0.67 MHz corresponding to the applied field of $B^{ext}$ =50 Oe. Another remarkable aspect concerns the ordering temperature of the Pr moments which appears to be strongly suppressed since $\nu_\mu$ does not exhibit a decrease below 20 K similar as in the pure Pr-1111 sample. This can be understood due to the non magnetic nature of the Sr ions which interrupt the magnetic exchange paths between the magnetic Pr ions. In return, this observation confirms that the magnetism in the FeAs layers cannot be explained in terms of a chemical segregation of the Sr ions such that the magnetic regions remain essentially undoped.

Our μSR data suggest that a mesoscopic phase segregation takes place in this hole doped sample which involves the formation of regions where the AF order is hardly perturbed and others that are non magnetic and most likely superconducting. Notably, this trend is fairly similar to the one that has been recently observed in the hole doped 122-type system $(Ba,Sr)_{1-x}K_xFe_2As_2$ [21-23]. This observation raises the question whether this segregation into sizeable regions that are either magnetic or superconducting is really caused by a corresponding modulation of the carrier density with regions that are essentially undoped and other that are highly hole doped. This suspicion arises because such a charge density modulation is opposed by strong Coulomb forces which can only be overcome if a corresponding modulation in the density of the dopant ions is realised . A prominent example of such a mesoscopic segregation of the dopant ions is the cuprate compound $LaCuO_{4+\delta}$ where a superstructure formation of the excess oxygen ions is a well established experimental fact [24-27]. In comparison, for the hole doped pnictide 122- [23] and 1111-compounds [51] at present there exists no experimental evidence for a superstructure formation or another kind of mesoscopic clustering of the dopant ions. Even though, one cannot entirely exclude such a possibility until more detailed structural investigations have been performed.

A recent x-ray diffraction study on series of similar $Nd_{1-x}Sr_xFeAsO$ samples with 0≤x≤0.2 reveals that the a-axis lattice parameters at room temperature increase continuously with the Sr content as expected due to the larger ionic radius of $Sr^{2+}$ (1.18 Å) as compared to $Nd^{3+}$ (0.98 Å) [51]. These measurements are thus fully consistent with a random substitution of the Sr dopant ions. A splitting of the structural Bragg peaks is observed here only at lower temperature where a tetragonal to orthorhombic structural transition occurs for all samples between 120-140 K [51]. This finding agrees very well with our observation that a predominant part of the sample volume of the x=0.2 sample undergoes a magnetic transition at $T_N \approx 130$K. It also confirms the trend of the undoped and the electron doped samples that a static AF state develops only in the orthorhombic phase [8,10,15,17].

It appears that besides the carrier doping the transition from the static SDW order to the superconducting state is strongly dependent on additional parameters like the structural distortions of the Fe-As layers and the subsequent magnetic exchange interactions. An important role of the structural parameters is also suggested by recent reports that AF order can be suppressed and superconductivity induced by either applying pressure to the undoped compound (Ba,Sr,Ca)Fe$_2$As$_2$ [28, 29] or by isoelectronic substitution of P for As in BaFe$_2$As$_{2-x}$P$_x$ [30]. Both experiments hint towards the Fe-As-Fe bond angle as an important parameter in determining the magnetic and superconducting properties of these new-pnictide HTSC. The general trend suggested by these experimental observations is that superconductivity appears as soon as the static AF order is suppressed either by carrier doping or by other means.

We conclude this section by commenting on the superconducting properties of the x=0.2 sample. Unfortunately, from our present TF-µSR data we did not obtain any direct evidence of a superconducting response within the non-magnetic regions of the sample. Below $T_c^{ons} \approx 15$ K we did not observe any significant enhancement of the relaxation rate due to the formation of a vortex lattice nor a diamagnetic shift. At present we can only speculate that this might be explained due to a large value of the magnetic penetration depth, λ, that may even exceed the spatial extent of the SC clusters (which is not known). Besides, we remark that the value of the TF relaxation rate in this sample was already fairly high in the normal state, possibly due to the stray field from the magnetic regions or likewise due to some inclusions of the FeAs impurity phase that are resolved in the x-ray diffraction data [35]. Due to these difficulties we refrain from attempts to deduce a lower limit of the low temperature value of the magnetic penetration depth and to put in relation to a so-called Uemura-plot [45].

**Summary**

We reported muon spin rotation (µSR) measurements which revealed some interesting parallels between the magnetic and superconducting properties of the electron doped 122-type system BaFe$_{2-x}$Co$_x$As$_2$ and the ones that were previously reported for the electron doped 1111-type SmFeAsO$_{1-x}$F$_x$ system [17, 18]. In particular, we found that strongly disordered static magnetism coexists with superconductivity in underdoped samples and even at optimum doping we observed a slowing down (or enhancement) of dynamic magnetic correlations below $T_c \approx 25$K. In hole-doped 1111-type Pr$_{1-x}$Sr$_x$FeAsO we observed instead a mesoscopic phase segregation into regions with nearly unperturbed AF order and others that are non magnetic and most likely superconducting. The observed trend resembles the one of hole-doped 122-type $Ba_{1-x}K_xFe_2As_2$ [21-23] and thus seems to be fairly common in these hole doped systems. Our results suggest that the phase diagram of the pnictide superconductors exhibits a strong asymmetry with respect to doping with electrons and holes.

**Acknowledgment:**

This work is supported by the Schweizer Nationalfonds (SNF) by grant 200020-119784, the NCCR program MANEP, project 122935 of the Indo Swiss Joint Research Program (ISJRP), the Deutsche Forschungsgemeinschaft (DFG) by grant BE2684/1-3 in FOR538 and the UK EPSRC. The work in China is supported by the Natural Science Foundation of China, the Ministry of Science and Technology of China (973 project No. 2006CB60100) and the Chinese Academy of Sciences.